\documentclass[%
 reprint,
 amsmath,amssymb,
 aps,
]{revtex4-2}

\usepackage{graphicx}
\usepackage{dcolumn}
\usepackage{bm}

\usepackage[colorlinks = true,
            linkcolor = blue,
            urlcolor  = blue,
            citecolor = blue]{hyperref}
\usepackage{float}

\begin{document}

\preprint{APS/123-QED}

\title{Machine learning predicts extreme events in ultrashort pulse lasers}

\author{Myriam Nonaka, Monica Agüero, Alejandro Hnilo}

%


\author{Marcelo Kovalsky}
\affiliation{
 Laboratorio de Láseres Sólidos. Centro de Investigaciones en Láseres y Aplicaciones ( DEILAP ) - Instituto de Investigaciones Científicas y Técnicas para la Defensa ( CITEDEF) - CONICET, Villa Martelli, Buenos Aires, Argentina.
}%


\date{\today}

\begin{abstract}
In this paper we present a nonlinear autoregressive  neural network with a hidden layer of 50 neurons, three delays and one output layer that accurately is capable of predict the appearence of extreme events in a Kerr lens mode locking Ti:Sapphire laser with ultrashort pulses. Extreme events are produced in the context of a chaotic atractor and with chirped pulses.  The prediction of this neural network  works well with experimental and theoretical time series of amplitude of laser pulses. When fed with experimental time series we have 95.45\% of hits and 6.67\% of false positives while using theoretical time series the network predicts 100\% of extreme events but the false positive rise to 23.33\%.

\end{abstract}

\maketitle


\section{introduction}

Rogue waves, giant waves coming from nowhere, were a myth, just a tale from returning mariners to the pub, until the first scientifically documented rogue wave, the Draupner wave, occurred in 1995 in the North Sea and reported in \cite{RW}.
A wave is rogue if its height exceeds, at least, two times the significant wave height $H_s$, which is defined as the average wave height among one third of the highest waves in a time series. If we define the abnormality index (AI) as the quotient between the wave height and $H_s$, a wave is rogue if AI is greater than 2 \cite{libroRW}.
It is worth emphasizing that, from this definition, the concept of rogue wave is relative, that is, it depends on the complete set under study.

Once confirmed at sea, phenomena analogous to rogue waves were found in various physical systems such as deep water waves, liquid Helium \cite{EE_helio}, nonlinear optics \cite{EE_Solli}, microwave cavities \cite{EE_micro}, and even in finance \cite{EE_fin}.

In lasers, the first report of optical rogue waves, or extreme events, as are also known, was in a Kerr lens mode locked (KLM)  Ti:Sapphire laser \cite{EE1}.
All these extreme events  possess the characteristic phenomenological features of sea rogue waves;
they are extremely large and apparently unpredictable, follow unusual
L-shaped statistics, with kurtosis greater than normal, occur in a nonlinear medium, 
and are temporally steep compared with typical events.
Although the parameters that characterize an optical
system are of course very different from those describing waves on
the open ocean, the extreme events generated in the two cases bear
similarities.
A remarkable one is the role played by modulation instabilities in the formation of extreme events both in ocean \cite{mod_ins} and KLM Ti:Sapphire laser \cite{PRA_tiza}. In this way the study of EE in Ti:Sa laser, beyond its intrinsic interest, it is presented as an ideal test bed for the study of marine rogue waves. 

Predicting extreme events has been a goal since its discovery. Several authors have shown regularities before and after an extreme event in lasers with modulation of the field polarization  \cite{cyrille} and Nd:Vanadate with saturable absorber \cite{PRA_2015}. These observations implies that the trajectory in phase space associated with an EE is confined to a relatively well-defined manifold. This fact, consequently, encourages the possibility of forecast them.  In the Ti:Sapphire laser EE follow certain regularity or quasiperiodicity that comes from the periodicity of the spatial part of the pulse in a cold cavity.

In this paper we present an autorregresive neural network that forecast one step ahead and is capable of predict the extreme events that occurr in a KLM Ti:Sapphire laser. In order to train, validate and test the neural network we employ both experimental and theoretical time series. In both cases the percentage of right guesses is grater than 90\%, false positives less than 20\% and without false negatives.The rest of the paper is organized as follows: in section II we briefly review the physics of KLM Ti:Sa laser as well as the map model that describes its dynamics. In section III the neural network and its performance are detailed. Finally the conclusion are presented in section IV.
%

\section{Extreme events in the KLM Ti:Sapphire laser}

\begin{figure*} [ht!]
\includegraphics[scale=.45]{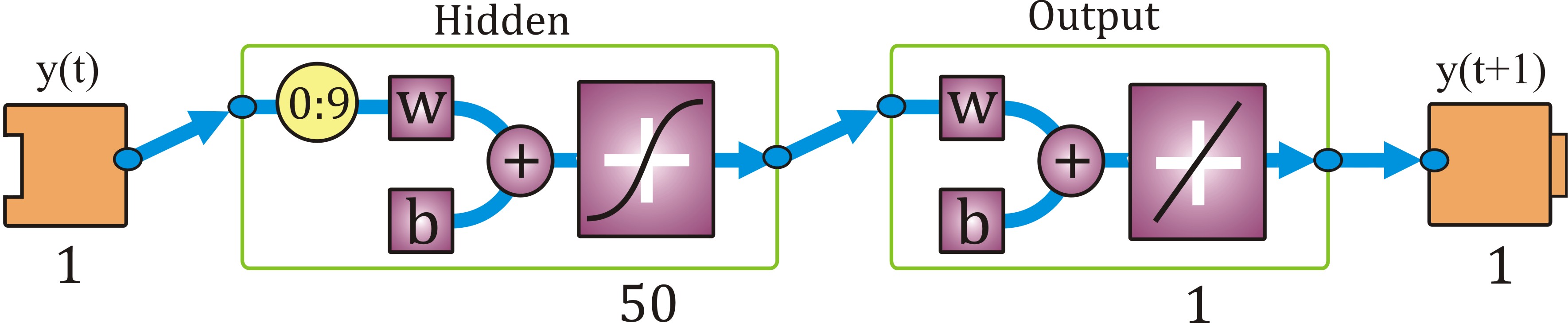}
\caption{Scheme of a NAR network with a sequence of values in a time series y(t) as input, one hidden layer with 50 neurons with 9 tap delays states as time delay line and a symmetric sigmoid activation function; the weight w and bias b; one output layer with linear activation function and y(t+1) as output prediction. }
\label{fig:a}
\end{figure*}

The Kerr lens mode locked Ti:Sapphire laser employed was a typical 7 elements, X configuration cavity, with a total length of 1724\;mm, which corresponds to a round trip frequency of 87\;MHz. For a 5\;W of cw pump at 532\;nm, an average power of 400\;mW in the spectral region of 800\;nm is obtained. This laser displays three coexistent  modes of operation: continuous wave, transform limited pulses, named P1 mode and positive chirped pulses, named P2 mode. For the chosen values of the control parameters, pulse duration for P1 mode is 35\;fs and for P2, 65\;fs.
The complexity of this laser is inherently intricate, as it operates within a delicate equilibrium of various spatial and temporal factors. In the temporal realm, the group velocity dispersion (GVD) across all optical components and the intensity-dependent self-phase modulation (SPM), predominantly occurring in the laser rod, find equilibrium through the counteracting negative dispersion generated by a pair of prisms within the cavity. In the spatial context, the dominant effects comes from the geometric configuration of the cavity and the intensity-dependent self-focusing (SF). Amplification in the active medium introduces additional nonlinearity through gain saturation. Both the SPM and SF effects intertwine to couple pulse energy, beam size, and pulse duration. A map model that take into account the nonlinearities and employ the spot size, the beam curvature radius,the
pulse duration, the chirp and the energy of the pulse as variables, accuratelly reproduce the dynamics of the laser. A detailed and complete description of the model, based in the five variable map, can be found in \cite{PRA_tiza}. 
  As the GVD of the laser cavity is adjusted close to zero from the negative side, the pulsed modes evolve towards
chaos following a different route: P1 through quasiperiodicity,
P2 through intermittency. Is important emphasize that EE appear only in  P2 mode of operation both in the theoretical model as well as experimentally.

\section{Prediction with neural networks}

We build a nonlinear autoregresive neural network (NAR) with the framework of MATLAB Deep Learning Toolbox. The key characteristic of a NAR neural network is its autoregressive structure. This means that the network uses its own past outputs as inputs to predict future values. In other words, it models the relationship between a sequence of its own past observations to make predictions. Unlike traditional autoregressive models, NAR neural networks incorporate nonlinear activation functions in their hidden layers. This allows the model to capture and learn complex, nonlinear relationships within the input data. NAR models typically consist of one or more hidden layers of neurons with nonlinear activation functions. The architecture may vary depending on the specific problem and the complexity of the data. NAR neural networks are trained using backpropagation, a supervised learning algorithm. The network is fed with input-output pairs, and the weights are adjusted iteratively to minimize the difference between the predicted and actual outputs. Our NAR has one input, a time series of the laser pulse intensity, one hidden layer of 50 neurons, three delays and an output layer, in that way, we use three elements to learn and predict the fourth. In Fig. \ref{fig:a} can be seen a graphic representation of the NAR that will be created and trained to predict a timestep ahead y(t+1) from past values of inputs y(t). 
There are no ``recipe'' or fixed rule in order to choose the optimal values of the neural network, so we use trial and error having in mind to improve the performance of the whole system. As learning criteria we use the mean square error (MSE) and two methods training were applied: Bayesian and Levenberg - Marquard. As activation function we employ tanh function. From the whole data available, we use 70 percent to training process, 15 percent of the data to validation and 15 percent to test the results. It is worth to mention that the selection of data in each stage is taken randomly and the weight w and bias b are calculated automatically from the training set by the neural network.

\begin{figure} [ht!]
\includegraphics[width=1\linewidth] {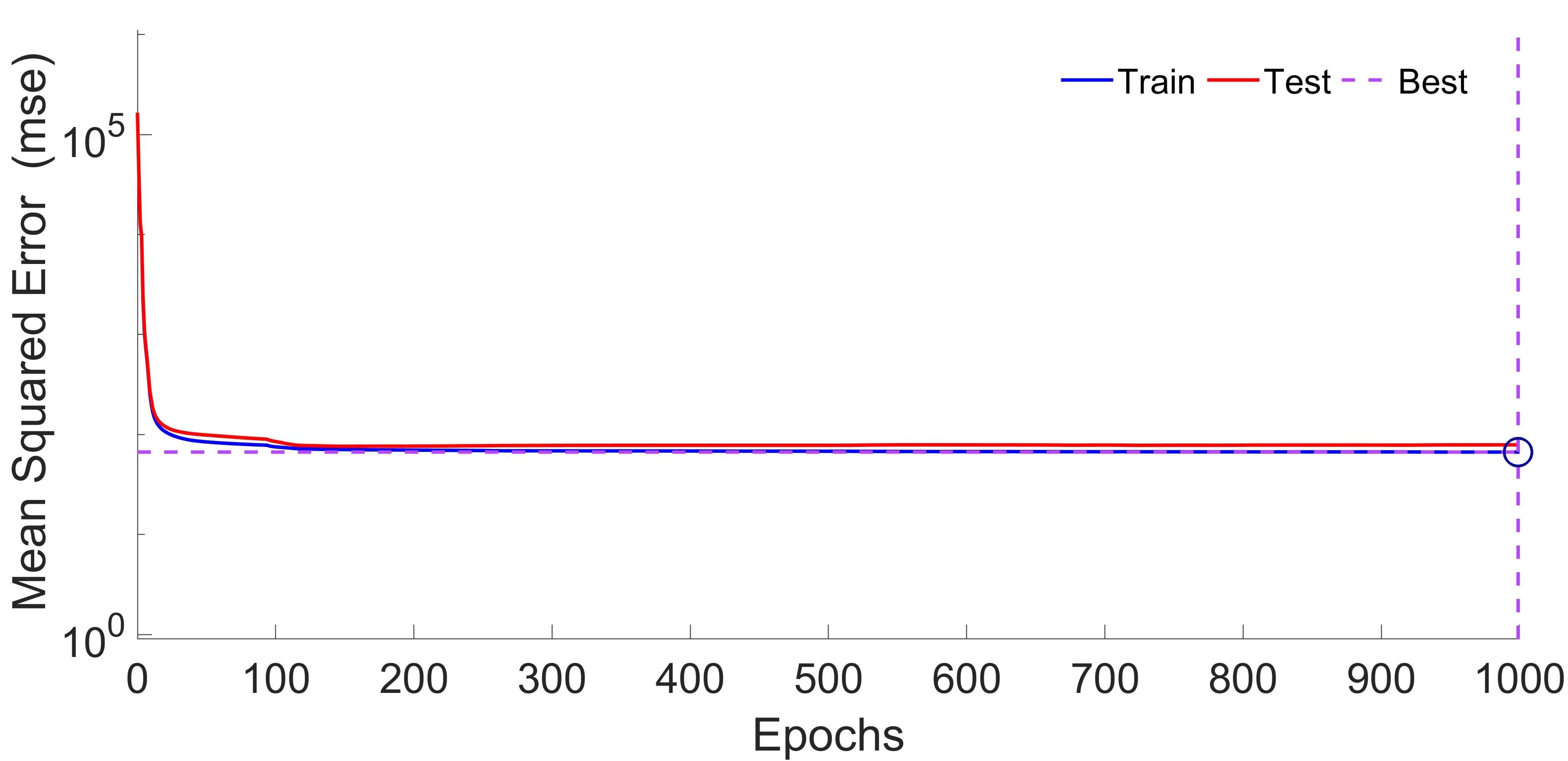}
\caption{Performance of NAR with experimental time series.The convergence between train and test up to 1000 epochs indicates that there is no overfitting.}
\label{fig:c}
\end{figure}

\begin{figure} [ht!]
\includegraphics[width=0.85\linewidth] {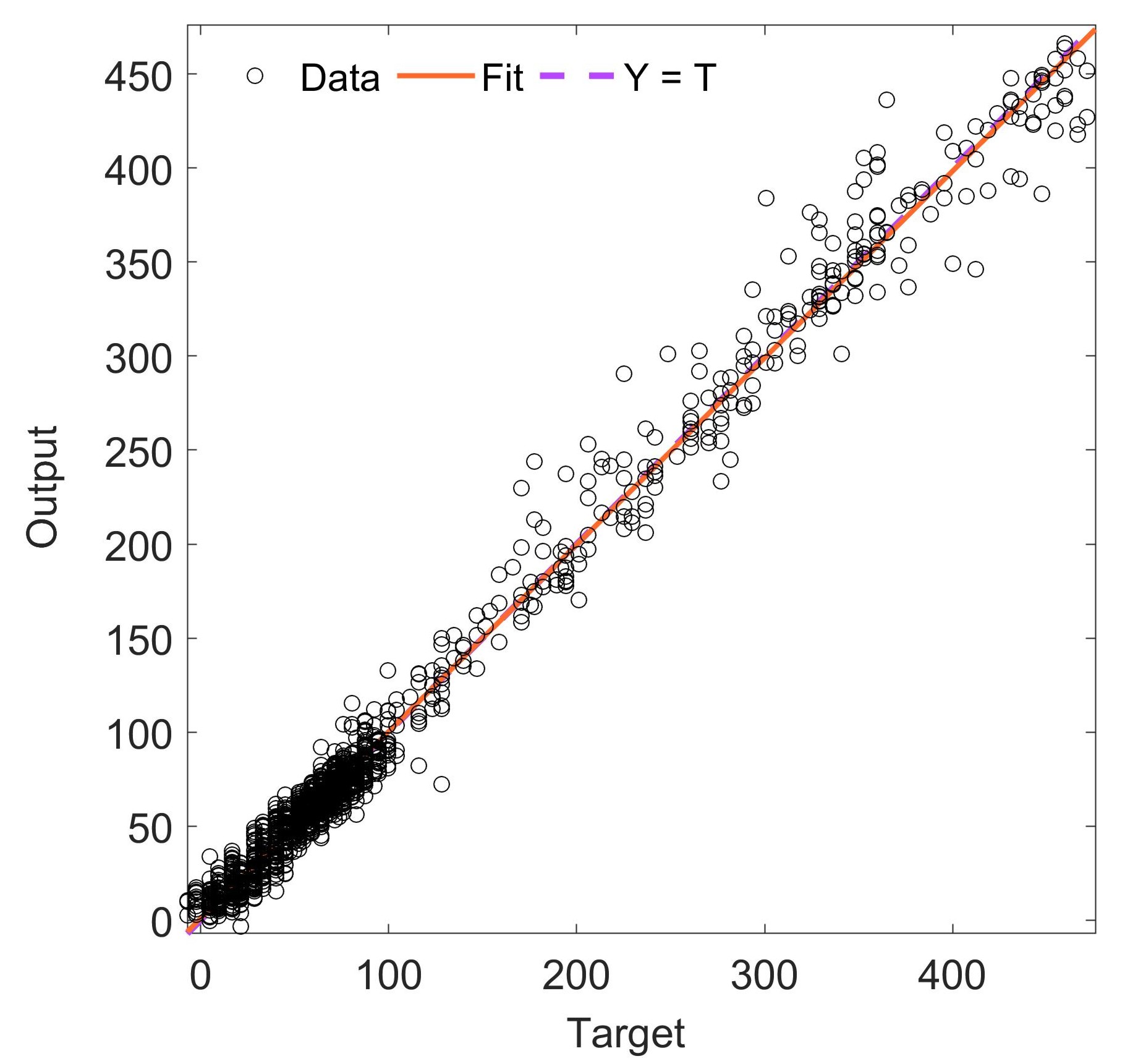}
\caption{Output (predicted values) vs. target (real values) from an experimental time series. Regression coefficient r=0.99298. Bayesian training. The line y=t overlaps with the fit line, so it is not visible in the graph.}
\label{fig:b}
\end{figure}

In that way, the parameters of the NAR were selected seeking a compromise between the best performance and computational speed. In order to quantify ``best performance''  we use as accuracy criterion MSE which is the standard deviation of the residuals, that is, prediction errors. Residuals are a measure of how far from the regression line data points are; MSE is a measure of how spread out these residuals are. In other words, it tells you how concentrated the data is around the line of best fit. Fig. \ref{fig:c} shows a typical learning curve of the performance of the neural network with an experimental time series. It is clear the convergence of training and validation curves as the NAR evolves in epochs ( an epoch is the number of passes a dataset takes around an algorithm) which indicates that the NAR is working well.  The other quantity used to check the performance of the prediction is the linear regression coefficient, r. When plotting predicted against real values, an exact prediction lies on a line with slope 1, so that r is a measure of the accuracy of prediction. Fig. \ref{fig:b} shows a typical r curve for our experimental data. We also seek, checking at each step these graphics, to avoid under and over fitting.

In order to predict with our NAR we employ two sets of time series, experimentals, from our Ti:Sa laser and simulated generated by our 5 equation map model. In both cases we apply two training methods, Bayesian and Levenberg - Marquard.

\begin{figure} [ht!]
\includegraphics[width=1\linewidth] {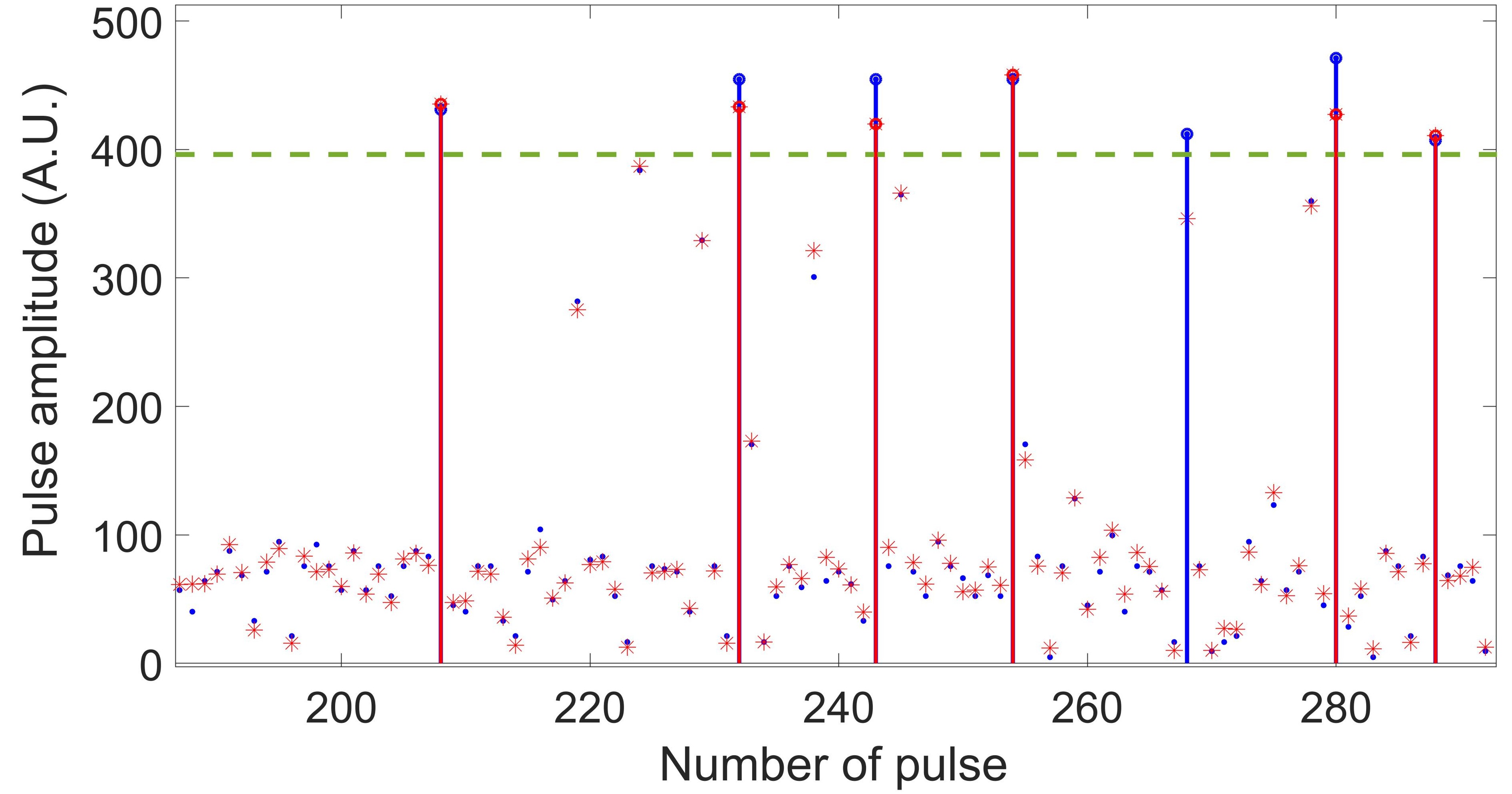}
\caption{ Extreme events: real vs. predicted for the experimental time series. X - axis: number of pulse, Y -axis: pulse amplitude (arbitrary units). Vertical red line indicates the EE threshold. Blue dots: amplitude of predicted laser pulse. Red dots: amplitude of real laser pulse.Those that exceed the threshold are EE and are marked with solid vertical lines. Blue for prediction, red for real.
  }
\label{fig:d}
\end{figure}

We set, for Bayesian training method, an input delay of 10 and employ an experimental time series of 10000 pulses. 
The test portion of the time series contains 44 extreme events. Our NAR predicts 45 EEs, 42 of which match the true ones, that is a 95.45 percent of real EEs and 3 that our model identifies,wrongly, as EE, that is false positives.
By changing to Levenberg Marquard training, we use an input delay of 9. In this case, the test portion of the time series has 46 EEs. Our NAR predicts 46 EEs, but only 43 are real ( 93.47\%) and 3 are false positives( 6.53\%). In Fig. \ref{fig:d} we plot a portion of the predicted and real time series. Red dots represent actual laser pulse amplitude while blue dots are predictions. The horizontal solid line mark the EE limit given by the abnormality index. Vertical lines indicate actual (red) and predicted (blue) extreme events. Notice that amplitude of predicted EE is slightly  greater than actual ones. In that way a large pulse, but below the AI limit, can be predicted as an EE, thus creating a false positive. Anyway, as is summarized in Table \ref{tab:tab1}, the percentage of false positive is low.

\begin{table}[ht!]
\caption{Prediction results: Results of prediction with an experimental time series trained with Bayesian and L-M methods. In both cases the number of hits, that is when the NAR predicts the next pulse as an EE in coincidence with the actual pulse, exceed ninety percent.}
\begin{ruledtabular}
\begin{tabular}{p{2cm}cddd}
\centering{Training Method}&No EE&
\multicolumn{1}{c}{\textrm{Pred EE}}&
\multicolumn{1}{c}{\textrm{\% Hits}}&
\multicolumn{1}{c}{\textrm{\% False Pos.}}\\
\hline
&&\mbox{}&\mbox{}&\mbox{}\\
\centering{Bayesian} &44& 45 & 95.45 & 6.67 \\
\hline
\centering{Levenverg - Marquard}
  &46   & 46 & 93.47 & 6.53 \\
\end{tabular}
\end{ruledtabular}
\label{tab:tab1}
\end{table}

\begin{table}[ht!]
\caption{Results of prediction with an theoretical ( model simulated) time series trained with Bayesian and L-M methods. The performance is similar to the predictions with experimental time series in hits, but there are an increase in false positives.}
\begin{ruledtabular}
\begin{tabular}{p{2cm}cddd}
\centering{Training Method}&No EE&
\multicolumn{1}{c}{\textrm{Pred EE}}&
\multicolumn{1}{c}{\textrm{\% Hits}}&
\multicolumn{1}{c}{\textrm{\% False Pos.}}\\
\hline
&&\mbox{}&\mbox{}&\mbox{}\\
\centering{Bayesian}&23& 30 & 100 & 23.33 \\
\hline
\centering{Levenverg - Marquard}
  &21
  & 31 & 100 & 32.25 \\
\end{tabular}
\end{ruledtabular}
\label{tab:tab2}
\end{table}

\begin{figure} [ht!]
\includegraphics[width=1\linewidth] {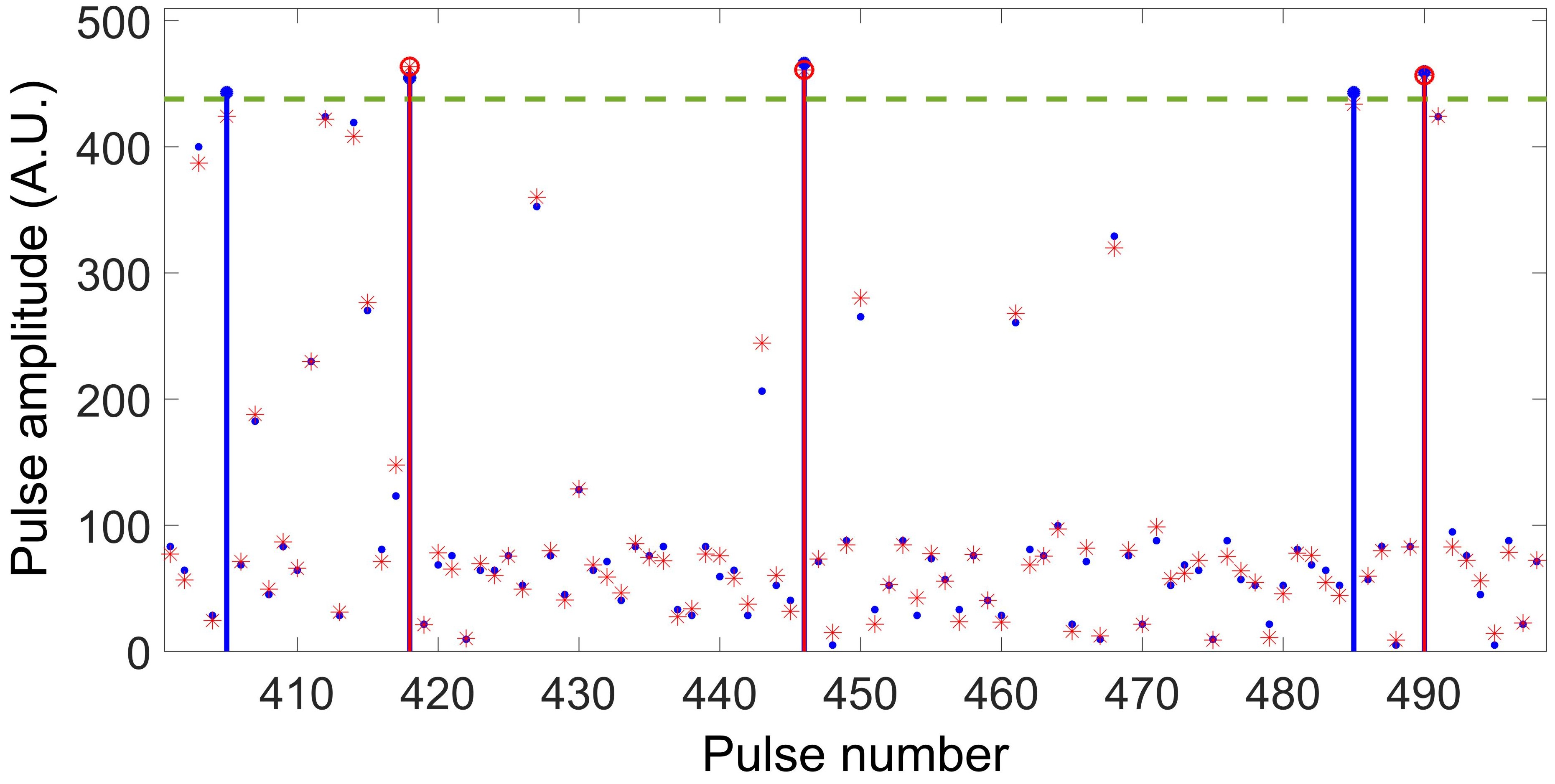}
\caption{Comparing extreme events: actual versus predicted values for the theoretical time series. X - axis: number of pulse, Y -axis: pulse amplitude (arbitrary units). The color-coding corresponds to that of Fig. \ref{fig:d}. It's worth noting that real pulses close to (but below) the threshold are predicted as extreme events, leading to increased false positives. }
\label{fig:e}
\end{figure}

If we use theoretical time series from the map model,instead, we find, with the Bayesian training and input delay of 10 that the NAR predicts successfully the existing 23 EEs but also forecast 7 false positives, as summarized in Table \ref{tab:tab2}. The amount of predicted false positives can be reduced by fine-tuning the AI limit value (remember that, in essence, this is an arbitrary value), that is, by adjusting the level of the vertical line in Fig.\;\ref{fig:e}.

\section{Conclusions}

    In this paper, we introduce a nonlinear autoregressive neural network architecture for predicting the occurrence of extreme events in a Kerr lens mode locking Ti:Sapphire laser with ultrashort pulses. The neural network consists of a hidden layer with 50 neurons, three delays, and an output layer. We predict one step ahead. The network demonstrates high accuracy in predicting extreme events, which are observed in the context of a chaotic attractor and with chirped pulses.

To evaluate the performance of the neural network, we conducted experiments using both experimental and theoretical time series data of the laser pulse amplitudes. When the network was trained and tested using experimental time series data, it achieved a hit rate of 95.45\% and a false positive rate of 6.67\%. 

Furthermore, we also tested the neural network using theoretical time series data. In this case, the network achieved a perfect prediction rate of 100\% for extreme events. However, the false positive rate increased to 23.33\%. This happens because our neural network has a small tendency to predict a larger pulse amplitude than real. In that way, pulses that are close, but below the EE threshold are predicted as EE. Since the abnormality index, that fixes the EE threshold, it is an arbitrary number, by fine tuning it, the rate of false positives can be easily lowered, without affect the predictions on the number of EE. 

These results demonstrate the effectiveness of the autoregressive neural network in predicting extreme events in the Ti:Sapphire laser system. The network's performance was particularly strong when trained and tested with experimental time series data, achieving a high hit rate and a low false positive rate. As a future improvement we will modify the neural network in order to predict more steps ahead, so that we can anticipate the ocurrence of an EE with more time.



\section{Funding}

Consejo Nacional de Investigaciones Científicas y Técnicas (CONICET) (PUE 229 20180100018 CO, PIP 2022 00484CO); Office of Naval Research Global (N62909-18-1-2021).

\section{Acknowledgements}

This work received support from the grants Office of Naval Research Global (U.S.), and CONICET (Argentina).

\nocite{*}


\end{document}